\documentclass[11pt,a4paper]{article}
\pdfoutput=1
\usepackage{jcappub}

\usepackage{color}
\usepackage{graphicx}
\usepackage[utf8]{inputenc}
\usepackage{amsmath, amssymb}
\usepackage{mathtools}
\usepackage{url}
\usepackage{wasysym}
\usepackage{float}
\usepackage{gensymb}
\usepackage{textcomp}
\usepackage{stmaryrd}
\usepackage{todonotes}
\usepackage{subfigure}

\newcommand{\mincir}{\raise
  -2.truept\hbox{\rlap{\hbox{$\sim$}}\raise5.truept \hbox{$<$}\ }}
\newcommand{\magcir}{\raise
  -2.truept\hbox{\rlap{\hbox{$\sim$}}\raise5.truept \hbox{$>$}\ }}
\newcommand{\siml}{\raise
  -2.truept\hbox{\rlap{\hbox{$\sim$}}\raise5.truept \hbox{$<$}\ }}
\newcommand{\simg}{\raise
  -2.truept\hbox{\rlap{\hbox{$\sim$}}\raise5.truept \hbox{$>$}\ }}

\begin{document}

\title{Zeldovich pancakes in observational data are cold}

\author[a,b]{Thejs Brinckmann,}
\author[b]{Mikkel Lindholmer,}
\author[b]{Steen Hansen,}
\author[b,c,d]{Martina Falco}

\affiliation[a]{Institute for Theoretical Particle Physics and Cosmology (TTK),
RWTH Aachen University, D-52056 Aachen, Germany}
\affiliation[b]{Dark Cosmology Centre, Niels Bohr Institute, University of Copenhagen, \\
Juliane Maries Vej 30, DK-2100 Copenhagen, Denmark}
\affiliation[c]{Dipartimento di Fisica, Universita' di Torino, \\
V. Pietro Giuria 1, I-10125 Torino, Italy}
\affiliation[d]{Istituto Nazionale di Fisica Nucleare (INFN), Sezione di Torino, \\
V. Pietro Giuria 1, I-10125 Torino, Italy}

\emailAdd{brinckmann@physik.rwth-aachen.de}
\emailAdd{wxp363@alumni.ku.dk}
\emailAdd{steen@dark-cosmology.dk}
\emailAdd{mafalco@unito.it}

\abstract{The present day universe consists of galaxies, galaxy clusters, one-dimensional filaments and two-dimensional sheets or pancakes, all of which combine to form the cosmic web. The so called "Zeldovich pancakes", are very difficult to observe, because their overdensity is only slightly greater than the average density of the universe.

\citet{Falco} presented a method to identify Zeldovich pancakes in observational data, and these were used as a tool for estimating the mass of galaxy clusters. Here we expand and refine that observational detection method. We study two pancakes on scales of 10 Mpc, identified from spectroscopically observed galaxies near the Coma cluster, and compare with twenty numerical pancakes.

We find that the observed structures have velocity dispersions of about 100 km/sec, which is relatively low compared to typical groups and filaments. These velocity dispersions are consistent with those found for the numerical pancakes. We also confirm that the identified structures are in fact two-dimensional structures. Finally, we estimate the stellar to total mass of the observational pancakes to be $2 \cdot 10^{-4}$, within one order of magnitude, which is smaller than that of clusters of galaxies.}
\keywords{cosmology: theory; cosmology: dark matter; galaxies: clusters:
  general; methods: numerical}

\maketitle

\section{Introduction}
\label{sec:Intro}
\begin{figure}
\centering
\includegraphics[width=0.75\textwidth]{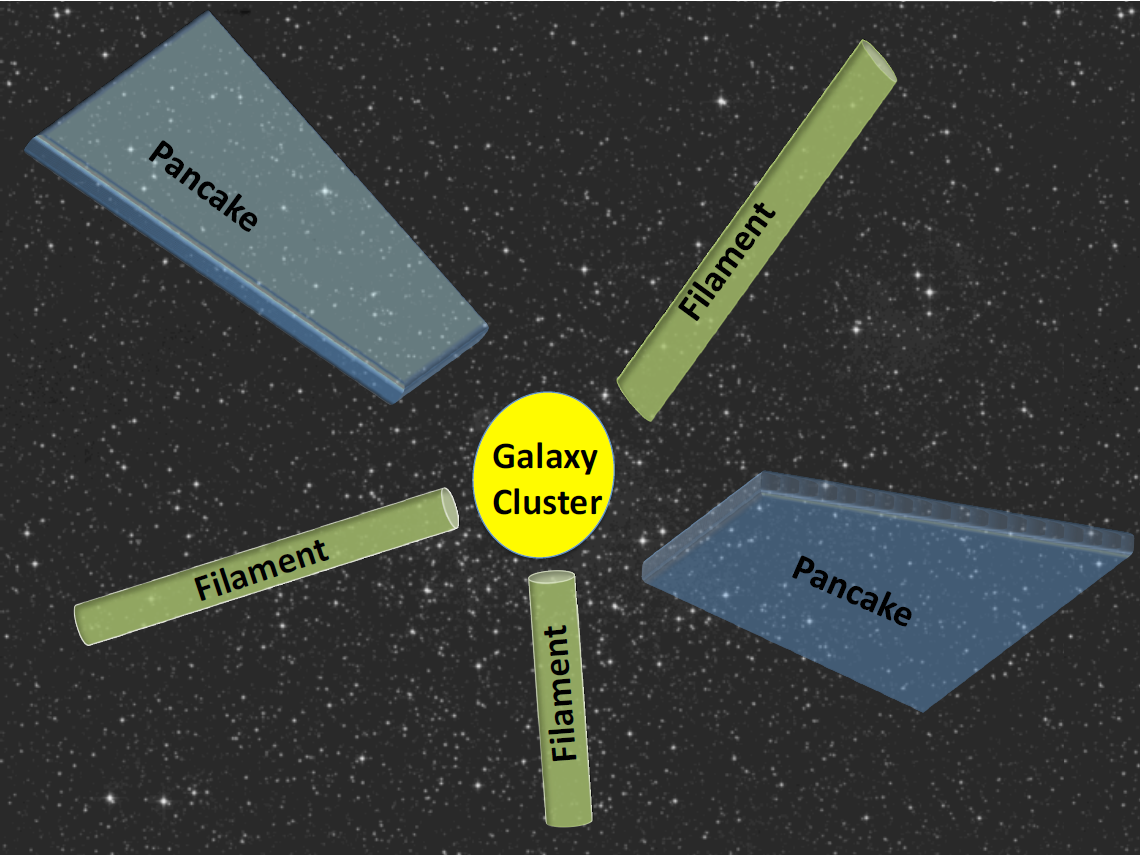}
\caption{The standard model of structure formation predicts that there should be dense galaxies and clusters, one-dimensional filaments, less dense two-dimensional pancakes, and underdense voids. In a hierarchical Universe all these structures have abundant substructures on all scales.}
\label{fig:pancakes}
\end{figure}
Our current understanding of the Universe, in the form of the standard cosmological model, predicts that there are a number of different types of structures to be found in the universe. Tiny perturbations in the matter density in the early universe will, under the influence of the force of gravity, evolve into various structures~\citep{Bond1996}. The early universe has an almost uniform three dimensional distribution, and a given perturbation on some scale will collapse first in one dimension and form two dimensional sheets (full of substructures), so called Zeldovich pancakes~\citep{Zeldovich1970}. These pancakes will then collapse in the other dimensions, forming filaments, galaxy clusters and groups.

As such, we would expect a universe filled with clusters of galaxies interlinked or fed by filaments and pancakes, not unlike the cosmic web seen from numerical simulations~\citep{Springel2005}. Galaxies and clusters of galaxies are easy to observe, but filaments, and especially pancakes, are more difficult to observe. The reason for this is, that they are considerably more diffuse and can be difficult to pick out from the average density of the Universe, as they are only slightly more dense~\citep{AragonCalvo2010a,Shandarin2012}.
	
\subsection{Finding pancakes}
\begin{figure}
	\centering
	\begin{subfigure}
		\centering
		\includegraphics[width=0.45\textwidth]{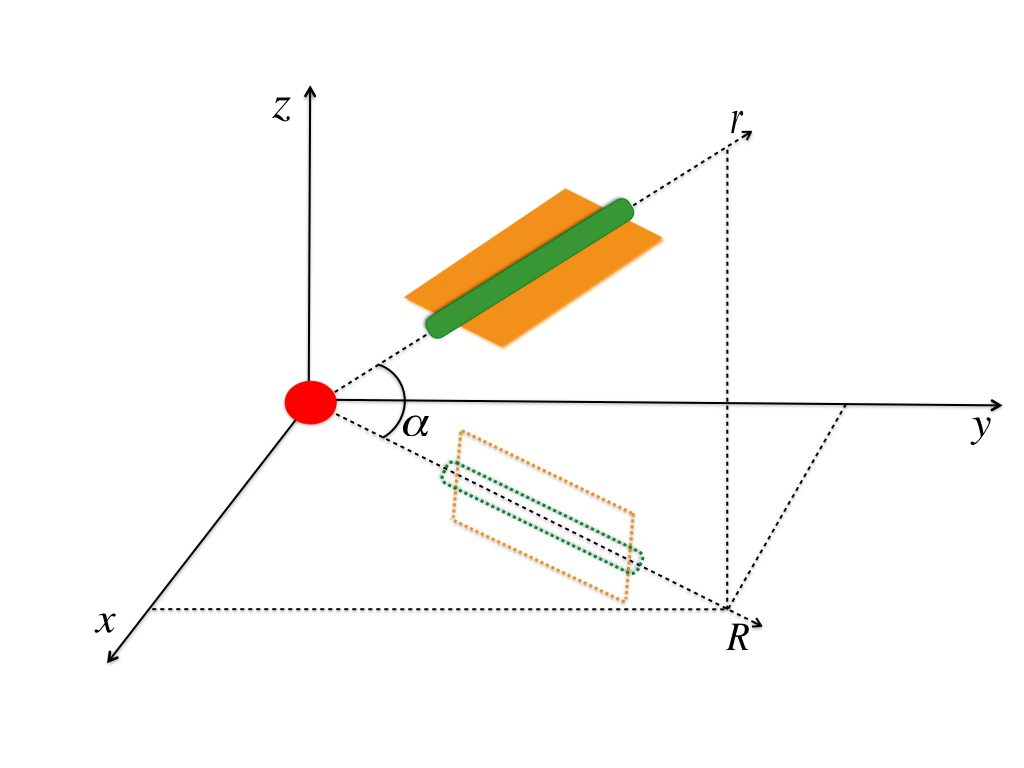}
	\end{subfigure}
	\begin{subfigure}
		\centering
		\includegraphics[width=0.45\textwidth]{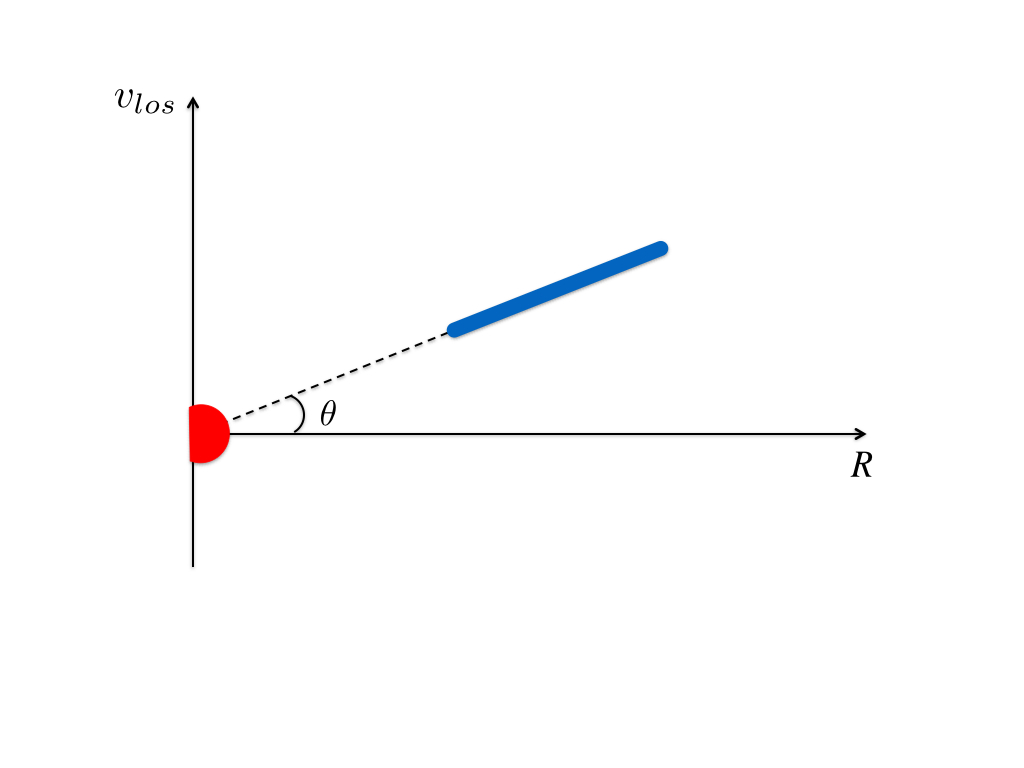}
	\end{subfigure}
	\caption{Schematic figures illustrating the observables. \textbf{Left}: We see the spatial coordinates x,y,z and the actual radius $\rm r=\sqrt{\rm x^2 + \rm y^2 + \rm z^2}$. The pancake is moving with a velocity v$_{\rm r}$ along the r-axis. In this case, x and y would be the coordinates on the sky and z would be the line of sight direction, so we would have the velocity in the line of sight direction given by $\rm v_{\rm los}=\rm v_{\rm z}=\sin(\alpha) \rm v_{\rm r}$. The radius when projected onto the (x,y)-plane would be $\rm R=\sqrt{\rm x^2 + \rm y^2}$ and the projection angle between the (x,y)-plane and the actual radius r is denoted by alpha. \textbf{Right}: A structure like this would look like a straight line in projected phase-space (line of sight velocity v$_{\rm los}$ vs projected radius R.), with a non-zero angle $\theta$.}
	\label{fig:projection}
\end{figure}
A pancake is a cosmological structure which is collapsing (or has collapsed) in only one dimension (see Figure \ref{fig:pancakes}). This means that a pancake seen face-on may appear almost uniform on the sky. If the pancake is near a massive galaxy cluster, then the motion of the pancake galaxy members due to the Hubble expansion will be perturbed by the galaxy cluster, leading to the pancake appearing as a relatively straight line in projected phase-space (R,v$_{\rm los}$) of projected radius and line of sight velocity~\citep{Falco} (see the schematic figure \ref{fig:projection}).

The clear infall lines in 3-dimensional radius and radial velocity space (r, v$_{\rm r}$), arise due to the coldness of the infalling matter in 3-dimensional space~\citep{Cuesta2008}. However, these lines entirely disappear when plotting the same data in projected phase-space, (R, v$_{\rm los}$). Therefore, smooth accretion leads to a fairly uniform distribution in projected phase-space. This is crucial, since observations of galaxy motion measure only projected positions on the sky and line-of-sight velocities. Only structures like groups, filaments and pancakes will lead to clear substructure in projected phase-space.

In this paper we discuss a method, for finding and identifying pancakes in observational data. This method uses only directly observable variables, namely the position on the sky (in the form of the projected radius) and the redshift of the galaxy. We will show the results of this method applied to the environs of the Coma galaxy cluster, outside the sphere of virialization, where no equilibrium has been achieved. Additionally, we will test if what we find indeed are pancakes, or if they are some other type of structure (see also \citet{WadekarHansen2014}).

\subsection{Other methods}
Many attempts have been made in the literature to identify or classify large scale structures, including pancakes, using a variety of methods, dating all the way back to~\citet{GellerHuchra1989}.

What most of these methods have in common, is that they are difficult to apply to observations and can only be used to identify structures in numerical simulations, due to three dimensional tensor and density field requirements. There are some exceptions~\citep{AragonCalvo2007,BondStraussCen2010,CostaDuarte2011}, which can be used on observational data to identify what \citet{Bond1996} refer to as walls, which are very large structures that consist of groups and clusters of galaxies, as well as filaments.

\citet{AragonCalvo2007} presented a method, the multiscale morphology filter (MMF), for identifying clusters, filaments and walls from samples of galaxies in redshift surveys or from cosmological N-body simulations. The method uses the Delauney Tessellation Field Estimator (DTFE) to create smoothed density fields, from which morphology response filters identify structures using Hessian matrix eigenvalues. It was successfully tested using Voronoi and N-body models, but can also be used to analyse observational data such as Fingers Of God from redshift surveys.

\citet{BondStraussCen2010} also use Hessian eigenvalues and eigenvectors of the smoothed density field to identify structures. With their algorithm they analyse mock galaxy distributions from a cosmological N-body simulation and observed galaxy distributions from the Sloan Digital Sky Survey (SDSS). They find similar morphology of structure for the two catalogs, with clumps dominating on smoothing scales of 5 h$^{-1}$ Mpc and filaments on scales of 10 to 20 or 25 h$^{-1}$ Mpc. Walls were not found to be dominant on the smoothing scales studied, which was up to 25 h$^{-1}$ Mpc, but the authors reported some evidence indicating a transition to wall-dominance at scales of 25 h$^{-1}$ Mpc.

\citet{CostaDuarte2011} found 212 or 444 pancakes (depending on selection criteria) and 204 or 436 filaments by analysing the density field of a sample from the SDSS DR7 using Minkowski Functionals to identify structures. In both observations and mock catalogues they found that filamentary structures tend to be richer, larger and more luminous than pancakes.\\

In addition to the methods that are capable of analysing observational redshift surveys, there is a large number of methods which can only be used to identify structures in cosmological simulations~\citep{Bond1996,Hahn2007,AragonCalvo2010b,Shandarin2012,Hoffman2012,Abel2012,Shandarin2014,Ramachandra2015,Falck2012,Falck2015,Vogelsberger2011}.

Some of these methods use dynamical properties~\citep{Bond1996,Hoffman2012,Shandarin2012}. \citet{Bond1996} use the eigenvalues of the velocity shear tensor to identify structures. \citet{Hoffman2012} defined what they call the V-web in terms of the eigenvalues of the velocity shear tensor and compared it to the T-web, which is defined by the eigenvalues of the Hessian of the gravitational potential, as used by~\citet{Hahn2007}. \citet{Hoffman2012} found the V-web to have increased resolution over the T-web. \citet{Shandarin2012} used a tessellation method to compute multi-stream and density fields and used Minkowski Functionals to identify structures. \citet{AragonCalvo2010b} created images akin to landscapes from the density field and used common image analysis methods to segment and analyse the images. They subsequently classified structures using the eigenvalues of the Hessian of the density field.

All these methods are successful at identifying structures in cosmological simulations, and are used to classify and study the different types of large scale structures of the cosmic web. Common amongst them is that they identify clumps, filaments and large wall-like structures from cosmological simulations, but cannot be used on observational data.

Additionally, there is a number of methods that use 6-dimensional phase-space to track dark matter streams and identify structures~\citep{Abel2012,Shandarin2014,Ramachandra2015,Falck2012,Falck2015,Vogelsberger2011}. Most of these methods track 3-dimensional dark matter manifolds in 6-dimensional phase-space, for example by using tessellation schemes to either directly track the 3-dimensional dark matter manifolds~\citep{Abel2012} or, equivalently, to track Lagrangian sub-manifolds~\citep{Shandarin2014,Ramachandra2015}, or by other similar methods~\citep{Falck2012,Falck2015}. Another related method for tracking dark matter streams is via integration of the geodesic deviation equation alongside the N-body equations of motion~\citep{Vogelsberger2011}. These methods have opened up new possibilities for studying the dynamical properties and non-linear clustering of dark matter structures in N-body simulations, but they cannot be used on observational data, where the full 6-dimensional phase-space information is not available.\\
\\

The paper is organized as follows. In section \ref{sec:Methods}, we present our method for pancake detection. In section \ref{sec:Numresults}, we use our method on simulated data, in order to check if the method finds pancakes. In section \ref{sec:Pancakes}, we show the results of applying our method to observational data of the Coma cluster and its environs. In section \ref{sec:Veldisp}, we test the coherency of the pancakes, as a confirmation that they indeed are pancakes. Finally, in section \ref{sec:Panmass} we obtain mass estimates of the pancakes.
\section{Method}
\label{sec:Methods}
\begin{figure}
\centering
\includegraphics[width=0.75\textwidth]{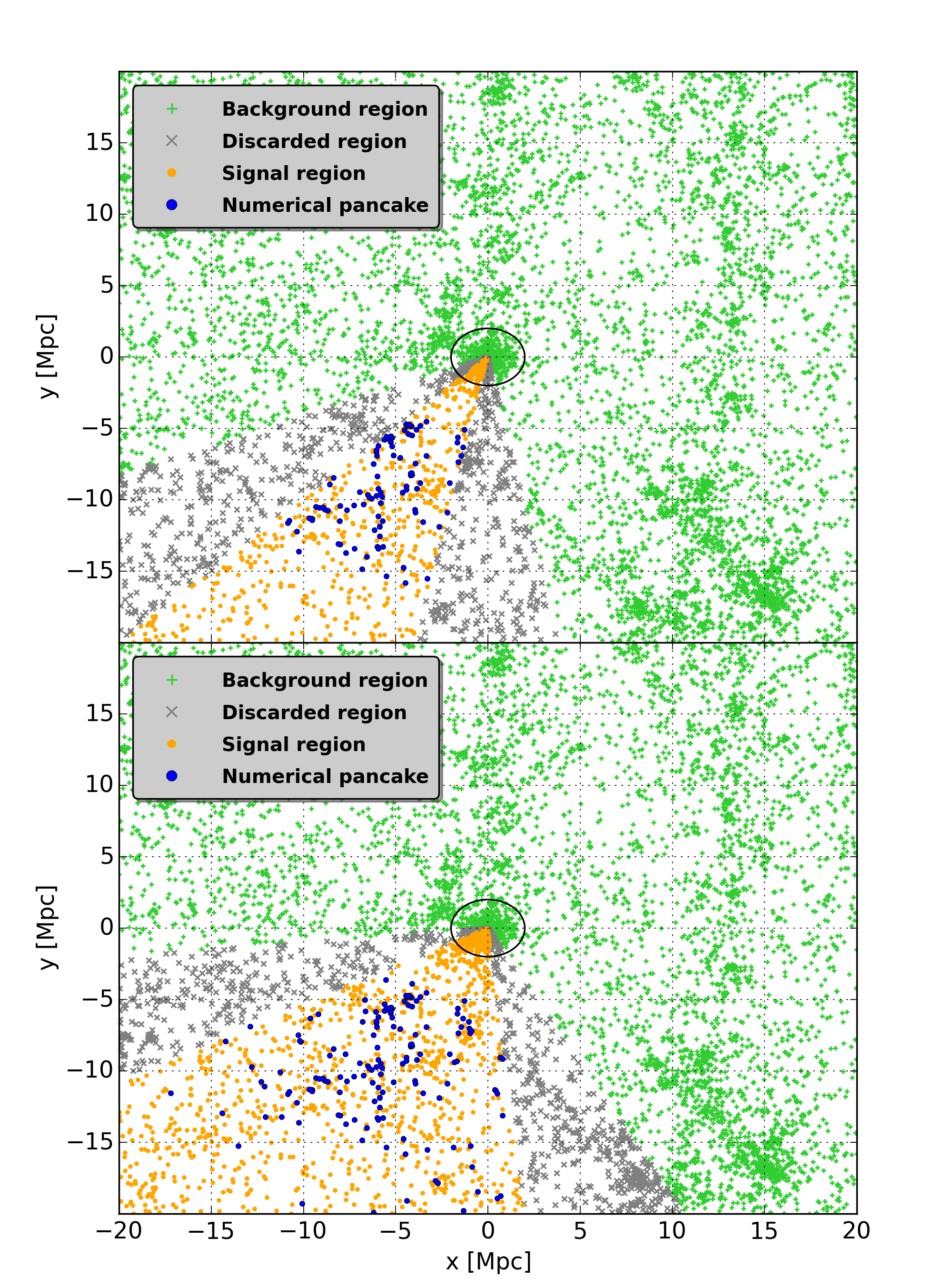}
\caption{A dark matter halo from a numerical simulation. \textbf{Top}: Spatial coordinates x-y mimicking right ascension and declination, as seen from observations. The background region is indicated by green crosses, the signal region by orange dots, the identified pancake by blue octagons and the discarded region as grey x's. The black circle indicates the virial radius of 2 Mpc. The pancake does not appear as a significant overdensity when viewed in x-y, as we would not expect it to. \textbf{Bottom}: Similar plot showing the effect of expanding the signal region.}
\label{fig:Numxy}
\end{figure}
Our objective is to identify pancakes in observational data. When we use the term pancakes, we are referring to structures with specific properties and not simply a two-dimensional overdensity on the sky. These pancakes are relatively cold and coherent structures, that have collapsed in only one dimension and have a very specific shape in radius-velocity space. Our search for these structures is independent of their appearance or overdensity on the sky, instead we identify them in projected radius-velocity space (R,v$_{\rm los}$), where they stand out as clear overdensities in the form of almost straight, inclined lines.

Groups of galaxies and filaments are clearly visible on the sky as over densities, whereas both have large velocity dispersions in projected phase-space. Only sheets with sufficiently fortunate orientation (face on) will appear as narrow lines in projected phasespace. Our method will therefore not identify all pancakes but only a fraction with the optimal orientation.\\
\\
The basic idea is to only use parameters which are directly observable, in such a way that the method can be applied to actual observational data.\\
\\
\textbf{Division of data into slices.}
We select a 35x35 Mpc$^2$ region and divide the cluster environs into 64 slices centred on the cluster in right ascension vs declination (RA,Dec) (see Figure \ref{fig:Numxy}).

We initially examine 4 adjacent slices at a time (this number is subsequently iterated upon), which is called the signal region. We discard four slices adjacent to the signal region, as this decreases the dependence of the detection on the choice of slices. Otherwise, the adjacent slices may interfere with the signal to background ratio. All other slices is considered the background region.

Pancakes are large, spatially diffuse objects. With current observational limits, determining the boundary of a pancake is difficult. Therefore, we define the boundary of a pancake as the edge of the selection of slices that optimize the signal compared to the background and noise. The choice of slices affect which galaxies are identified as belonging to the pancake, but the bulk of the structure remains the same. In general, with any additional slice added to the signal region, the number of galaxies incorrectly assumed to belong to the pancake region increases (e.g. galaxies that are in the pancake region in phase space, but are not nearby in real space). The objective, therefore, is to identify the bulk of the pancake, while minizimizing the noise and optimizing the signal compared to the background, rather than attempting to include every single galaxy belonging to the pancake.\\
\\
\begin{figure}
\centering
\includegraphics[width=0.75\textwidth]{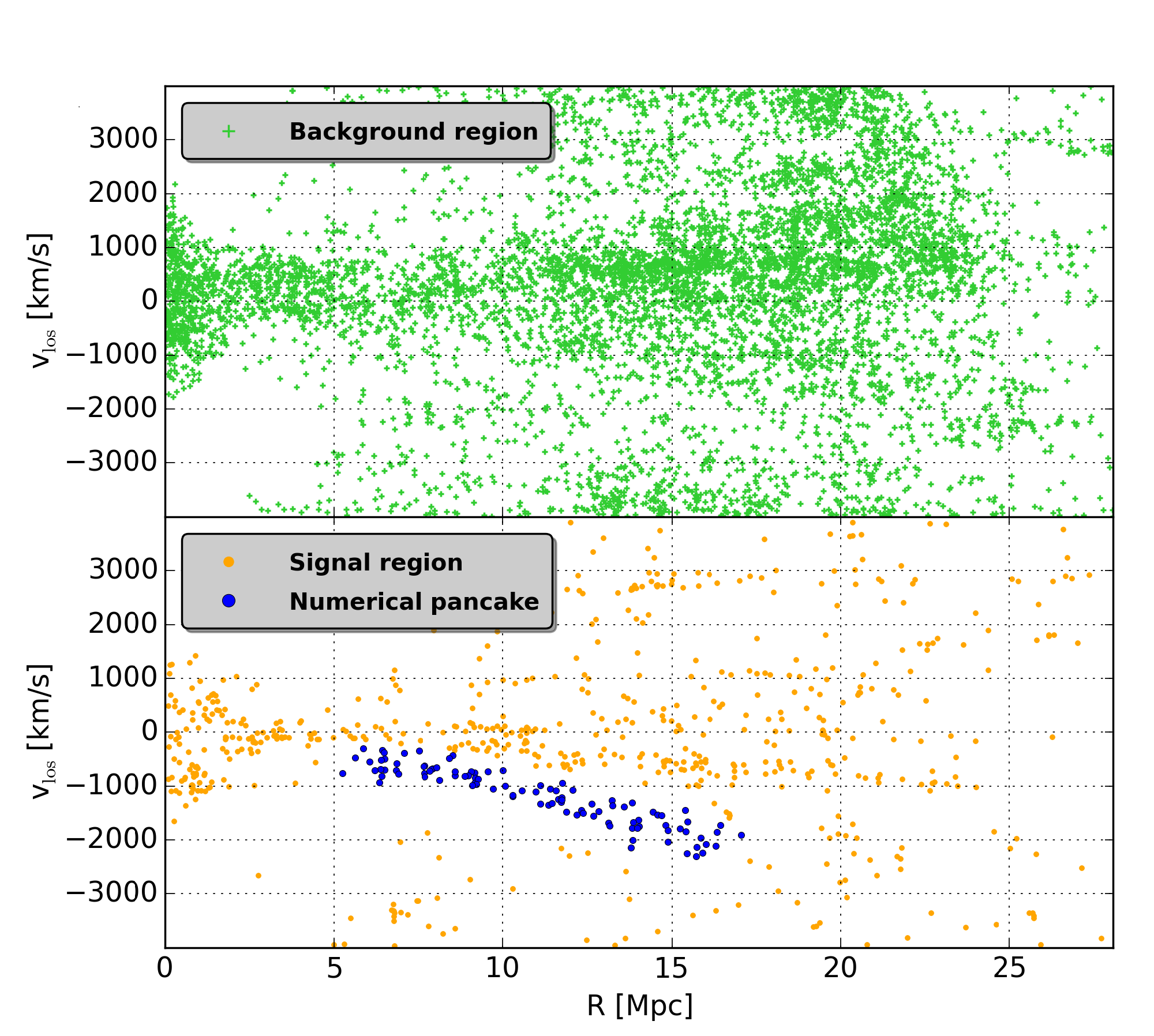}
\caption{Line of sight velocity as a function of radius for the mock catalogue. \textbf{Top}: Shows the background region indicated by green crosses. \textbf{Bottom}: Displays the signal region as orange dots along with the numerical pancake as blue octagons. We see that the pancake appears as a roughly straight inclined line, as expected. A hint of another straight inclined line is seen in the figure, but the structure is not selected by our method due to the imposed overdensity threshold and the low projection angle.}
\label{fig:NumRv}
\end{figure}
\noindent \textbf{Find overdensities.}
We look for pancakes in radius vs line of sight velocity (R,v$_{\rm los}$) by looking for overdensities (see Figure \ref{fig:NumRv}). As we are only interested in the non-virialized region, where the perturbation to their motion is almost linear, we remove the inner 5 Mpc in radius from the cluster centre.

We compare the density in the selected slices to the background density in (R,v$_{\rm los}$). We do this by counting the number of galaxies within an ellipse around each galaxy and comparing this to the number of galaxies within similar ellipses in the background, normalized by the background region to signal region ratio. We keep the signal data points that are above a certain threshold, where the threshold is an interval of 0.5 to 2 times the upper 2$\sigma$ value of the background number density distribution.

For each selection of slices we test a series of values in the interval and select the threshold that optimizes the signal compared to the noise and the background. The choice of threshold does not affect the resultant pancake significantly (as all galaxies in the pancake area in phase space are selected), but some pancakes may only be detected when using a lower overdensity threshold.\\
\\
\textbf{Selection of pancakes.}
If a pancake is present, and in our sensitivity range, we will obtain a rough pancake structure along with some noise. We iterate through the remaining data with our own friends-of-friends (FoF) algorithm and select the largest structure, which is usually the pancake, if one is present. However, occassionally a large amount of noise will be present in one area, such as around v$_{\rm los} \sim 0$.

Once we have selected the pancake we include additional galaxies along the edges of the pancake in projected phase space (R,v$_{\rm los}$), within a small ellipse around each galaxy in the pancake. This is in order to compensate for the selection bias in our method, which is biased against the lower density areas at the edges of pancakes.

At last we maximize the pancake signal to background ratio by adjusting the number of slices. We find that the optimal number of slices is typically 3 to 6.\\

This method is only able to find pancakes when their orientation is approximately face on, as our method has difficulties when the pancake is seen edge on. While seeing the pancake edge on will make it appear slightly more dense on the sky (but not sufficiently dense to pick out as a spatial overdensity), it would also mean the pancake spans several Mpc in the line of sight direction. In this case, the Hubble flow will disperse the pancake in in the line of sight direction in projected phase-space (R,v$_{\rm los}$), making it appear as diffuse straight inclined lines. These lines are not much more dense than the background, and, as such, are difficult for our method to detect. We estimate that our method is sensitive to an edge-on/face-on tilt of up to approximately $\pm 45$ degrees away from face-on.

Another limitation of our method is, that it is most sensitive to pancakes with an inclination relative to our line of sight (the angle $\alpha$ in Figure \ref{fig:projection}) in the range $\alpha = \pm$ 20 to 70 degrees. Pancakes outside of this interval will either be lost in the noise around v$_{\rm los} = 0$ in (R,v$_{\rm los}$) for small angles, or have little extent along the radius axis for large angles.\\

Our definition of a pancake is a structure that is collapsing (or has collapsed) in only one dimension. It is therefore in some degree of equilibrium in this dimension, but not in the other two dimensions, which are still in early stages of equilibration. As such, it is a two-dimensional structure that is relatively cold compared to groups and filaments. The structures we find are on scale 10 Mpc and are somewhat bulky structures with moderate axis ratios (see section \ref{sec:Numdist}). Visually on the sky they consist of solitary galaxies, interspersed with a few groups of galaxies.
\section{Finding pancakes in a mock catalogue}
\label{sec:Numresults}
We test our method on a cluster sized halo of pure dark matter particles from a cosmological N-body numerical simulation of pure dark matter based on WMAP3 cosmology, with cosmological parameters $\Omega_M=0.24$ and $\Omega_{\Lambda}=0.76$, and Hubble parameter $h=0.73$. The simulation consists of $1024^3$ particles, each of mass $2.554 \cdot 10^8 M_{\astrosun}$, in a box of size $160h^{-1}$ Mpc. Using the MPI version of the ART code~\citep{Kravtsov1997} \citep{Gottlober2008}, the behavior of the particles is simulated from redshift $z=30$. A hierarchical friends-of-friends algorithm with linking length of 0.17 times the mean interparticle distance was used to identify clusters and overdensities.

Twenty-five halos at redshift $z=0$, with virial masses in the range $M_{\rm vir}=(1.011-20.14) \cdot 10^{14} M_{\astrosun}$, virial radii $r_{\rm vir} = 1.2-3.3$ Mpc and virial velocities $V_{\rm vir} = 601-1630$ km/s were selected. In order to provide datasets of manageable size, 10000 dark matter particles, in regions of 20 x 20 Mpc$^2$ to 30 x 30 Mpc$^2$ with velocities of $\pm 4000$ km/s, were randomly selected. We only include a limited number of particles, in order to realistically optimize our method to observational data, where the number of galaxies in a given observed cluster is fairly small. The data is centered on the position of the most massive substructure found at a linking length 8 times shorter.

In the following we treat the dark matter particles as the galaxies from observations, which is a reasonable first approximation when examining dark matter dominated structures near galaxy cluster sized halos.\\

We apply our method to the 25 numerical halos, in the form of an automated overdensity finder in phase-space. From this we obtain a raw sample which includes pancakes and groups of galaxies. Since the groups of galaxies have roughly constant velocity across radius in projected phase-space, they are easily identified. In this way we finally select 20 pancake candidates.

In Figures \ref{fig:Numxy} and \ref{fig:NumRv} we show a sample pancake identified near a halo of virial mass $M_{\rm vir}=4.75 \cdot 10^{14} M_{\astrosun}$, virial radius $r_{\rm vir} = 2.0$ Mpc and virial velocity $V_{\rm vir} = 1007$ km/s. This is the only pancake located near this halo. Both our method and a similar method employed by \citet{Falco} find the same pancake near this halo.

For this numerical pancake we elected to keep 6 spatial slices, because reducing this number by any more slices lowers the signal considerably. Keeping the 6 slices posed another problem, in that there was a large amount of noise present around $v_{\rm los} \approx 0$. Our numerical pancake was identified by our FoF algorithm as being attached to a large almost horizontal strip, which we removed manually. Although the almost horizontal strip could appear to be another pancake, we did not identify it as one, since our method has limitations with regards to low projection angles, where the background is large. We also cut the inner 5 Mpc of the data, corresponding to roughly 2 virial radii.\\

In the following we will examine, whether the numerical pancakes have the spatial distribution of particles we expect of pancakes.

\begin{figure}
\centering
\includegraphics[width=84mm]{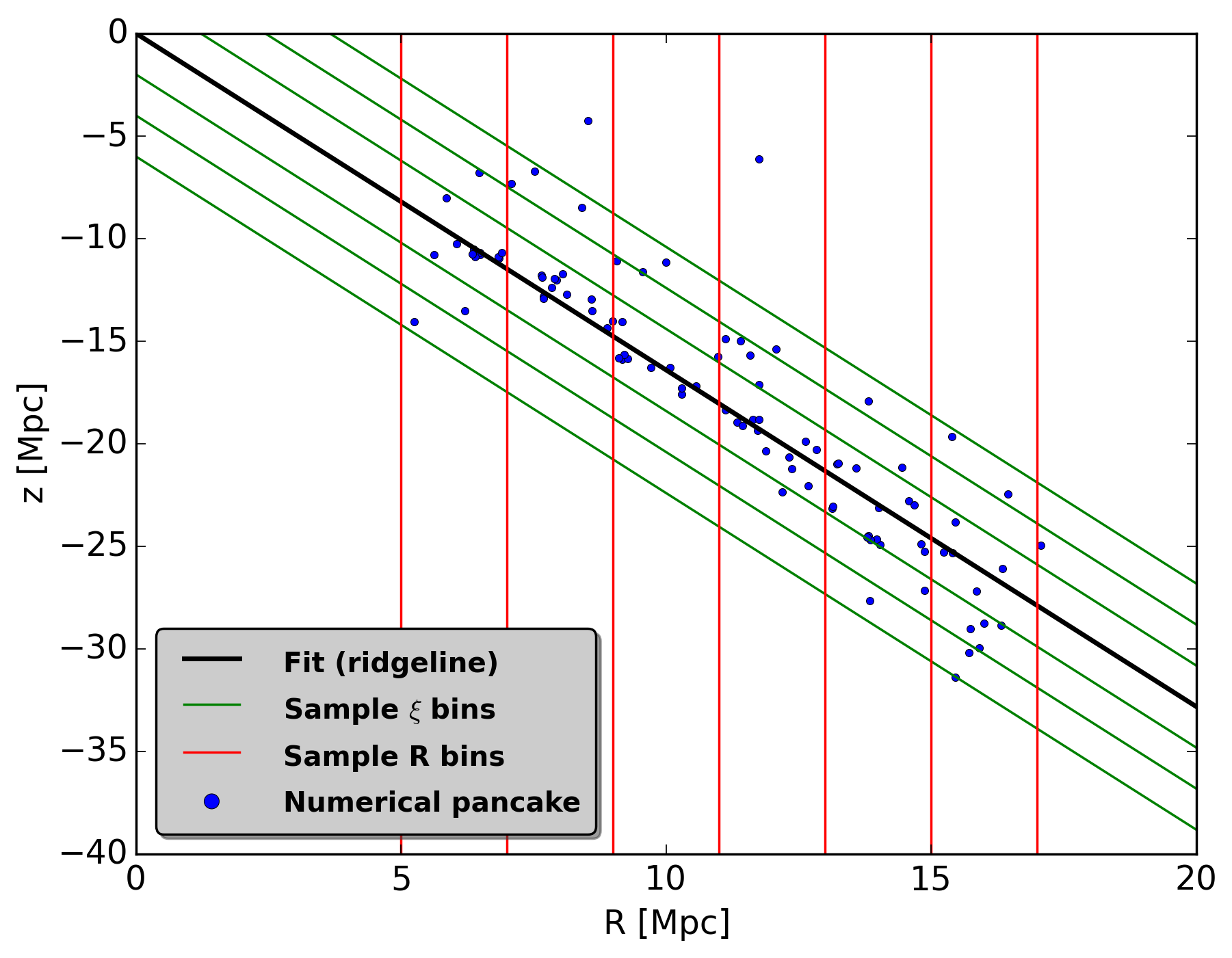}
\caption{Sample bins of the kind used in section \ref{sec:Numdist}. The figure shows radius, z distance from the halo centre (R,z) for the numerical pancake (blue octagons). The fit, or ridgeline, is shown as a black line. Sample bins along the radius axis are indicated by red lines and sample bins along the $\xi$ direction (the direction perpendicular to the fit) are seen as green lines.}
\label{fig:samplebins}
\end{figure}

\subsection{Numerical pancake particle distribution}
\label{sec:Numdist}
The galaxy distribution of an equilibrated galaxy group or cluster will appear very different from a top-hat. The projected profile may be a Sersic or a broken power-law, and with low resolution it may resemble a Gaussian distribution in all dimensions. Similarly, the collapsed dimension of a pancake is expected to resemble a Gaussian, and not a top-hat. If we look at the density profile of structures in (x,y) and (R,z), where R is the projected radius and x,y,z are the spatial coordinates, we would expect to see them collapsed in a different number of dimensions, depending on the structure. A collapsed dimension manifests itself as a roughly Gaussian distribution, when binning and plotting density of data points as a function of distance in bin direction. A non-collapsed dimension, however, will show roughly as a ``top hat'' shape in a similar plot. To reiterate, for pancakes we would expect only one collapsed dimension.\\

We can only conduct this test for the numerical simulation, where we are able to examine all dimensions. For observations we lack the spatial position along the line of sight for individual galaxies.\\

\begin{figure}
\centering
\begin{subfigure}
\centering
\includegraphics[width=0.3\textwidth]{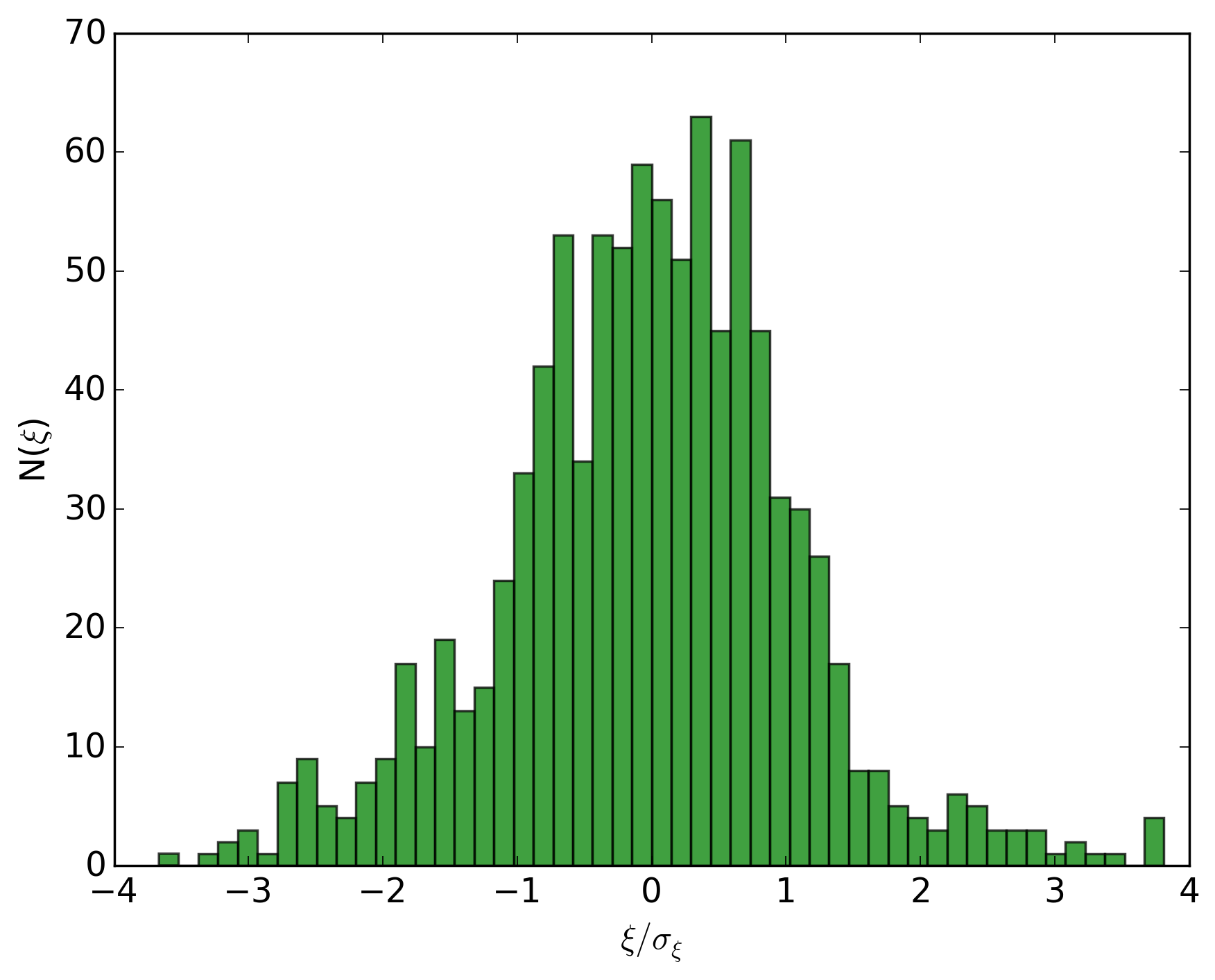}
\end{subfigure}
\begin{subfigure}
\centering
\includegraphics[width=0.3\textwidth]{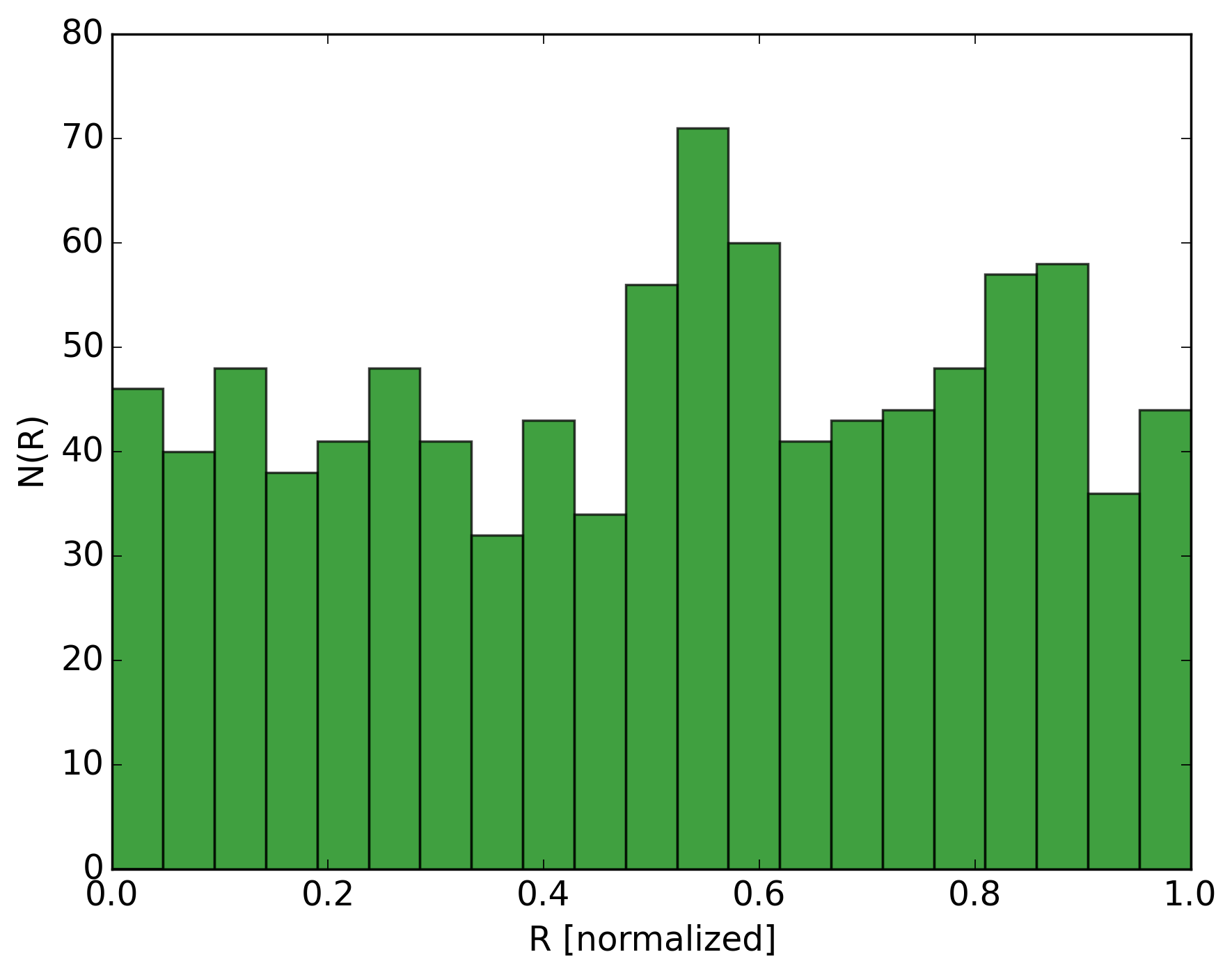}
\end{subfigure}
\begin{subfigure} %84mm}
\centering
\includegraphics[width=0.3\textwidth]{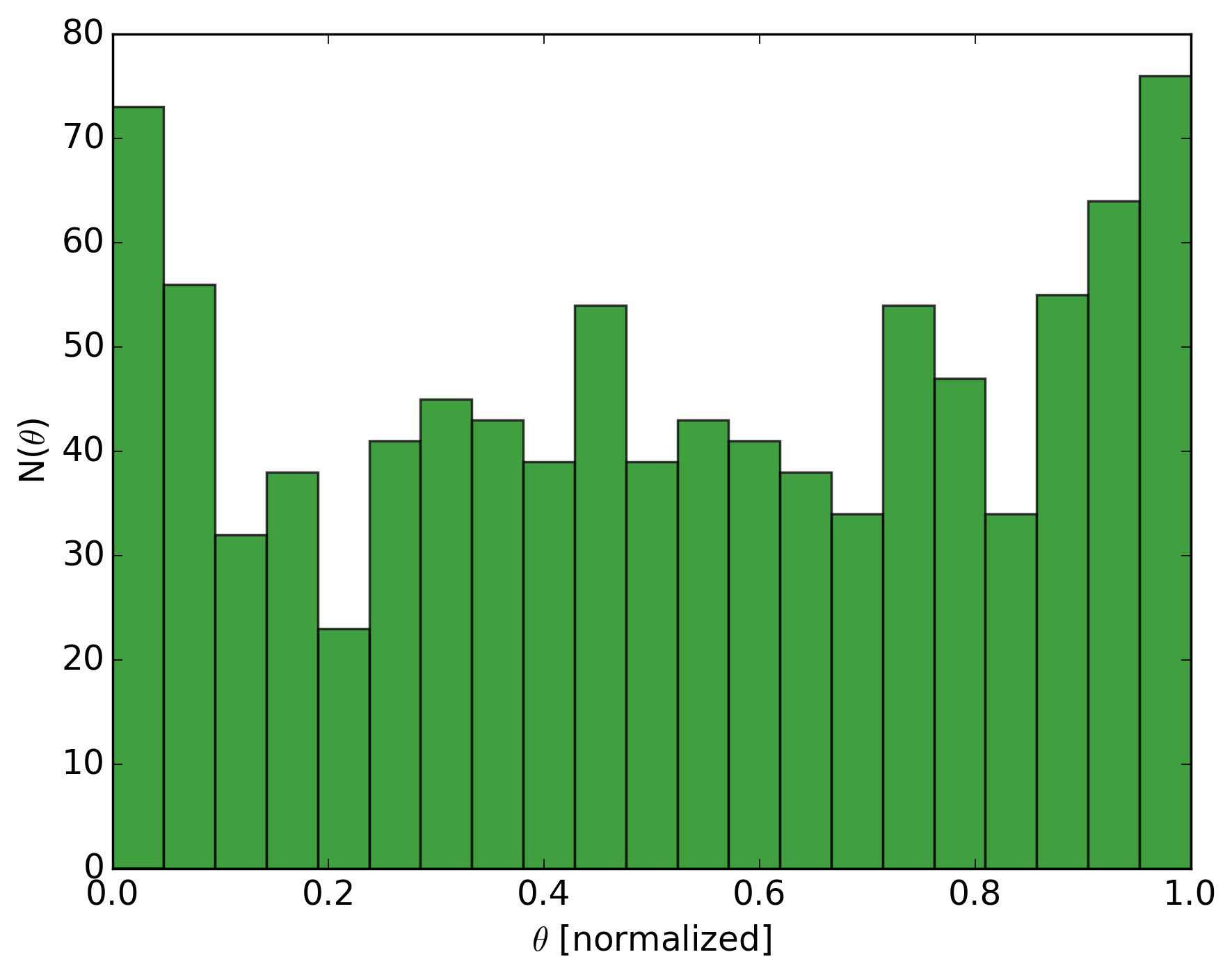}
\end{subfigure}
\caption{These figures show the distribution of galaxies along different directions for all twenty numerical pancakes combined. \textbf{Left}: Number of galaxies, N$(\xi)$, as a function of distance perpendicular to the pancake, $\xi$, in R,z (projected radius, line of sight distance), divided by $\sigma_\xi$. Here $\sigma_\xi$ for each pancake was used for that data. It displays a near Gaussian distribution, implying that the pancake is collapsing (or already has collapsed) in this dimension. \textbf{Center}: Number of galaxies, N(R), as a function of normalized projected radius, R, based on an R,z plot. The distribution of galaxies is roughly constant along this axis, as such the pancake is most likely not in any advanced stage of collapse in this dimension. \textbf{Right}: Number of galaxies, N$(\theta)$, as a function of aperture angle with centre in (0,0), $\theta$, based on an x,y plot (where x,y is the plane perpendicular to the line of sight along which a given pancake was identified). The distribution is also roughly constant along this axis, therefore the pancakes also have yet to collapse in this dimension.}
\label{fig:tophat}
\end{figure}

If we examine the line of sight direction, taken as the z direction, by looking at the z coordinate as a function of the projected radius (R,z), then we can bin in two different ways to extract information about two of the dimensions of the pancake (see Figure \ref{fig:samplebins}). First we fit the pancake with a straight line in (R,z), this line can be considered the one dimensional centre, or ridgeline, of the pancake. We then bin parallel to this line, which provides us with information about the thickness of the pancake and the distribution of galaxies along this dimension. For the twenty numerical pancakes the resulting combined distribution is given by Figure \ref{fig:tophat} (left), where we normalized the distance perpendicular to the pancake $\xi$, with the dispersion in that direction $\sigma_{\xi}$ for a given pancake. The plot shows number count per bin, N$(\xi)$, as a function of normalized distance perpendicular to the ridgeline of the pancake, $\xi/\sigma_{\xi}$. Here we see a rough Gaussian distribution, indicating a collapsing (or collapsed) dimension.

Next we bin perpendicular to the R axis, which tells us how the galaxies are distributed in a second dimension, namely along the length of the pancake. The resulting combined distribution for the twenty numerical pancakes is shown in Figure \ref{fig:tophat} (center), where we have normalized projected radius R vs number count per bin N(R). The projected radius was scaled by the minimum radius and normalized by the radial length of a given pancake. We see no hint of a Gaussian distribution, but rather something roughly resembling a top hat shape, indicating a non-collapsed dimension.

For the third dimension we examine the plane of the pancake face on, that is we take the x and y as positional coordinates on the sky. In (x,y) we bin along the aperture angle from the cluster centre, similar to the 64 spatial slices used in Figure \ref{fig:Numxy}. From this we can see the width of the pancake and the distribution of galaxies in this last dimension. The resulting combined distribution  for the twenty numerical pancakes is shown in Figure \ref{fig:tophat} (right). Here we have the normalized angle around a circle centred on the cluster, $\theta$ as x axis and the number count per bin, $N(\theta)$, as y axis. The $\theta$ values were scaled to start at zero and were normalized by the extent in the $\theta$ direction of a given pancake. Again we see no hint of a Gaussian distribution.\\

We see clearly, that the twenty numerical pancakes have one collapsed dimension and two non-collapsed dimensions, in accordance with our definition of a pancake.

Additionally, from the information in this section, we can examine the shape of the pancakes. If we consider the pancakes as tri-axial ellipsoids and take the radial extent as the major axis (c) and two times the one sigma value of the depth ($\sigma_{\xi}$) as the minor axis (a), we obtain a measure of the shape of the pancakes, $a/c = 0.4 \pm 0.2$. So we see that the pancakes are quite bulky structures, with moderate axis ratios. It should be noted, that our method for pancake selection makes this result conservative. Our pancakes could be longer (or wider), depending on selection criteria, but we include the full depth of the pancakes within the selected region.
\section{Two pancakes near Coma}
\label{sec:Pancakes}
\begin{figure}
\centering
\includegraphics[width=0.75\textwidth]{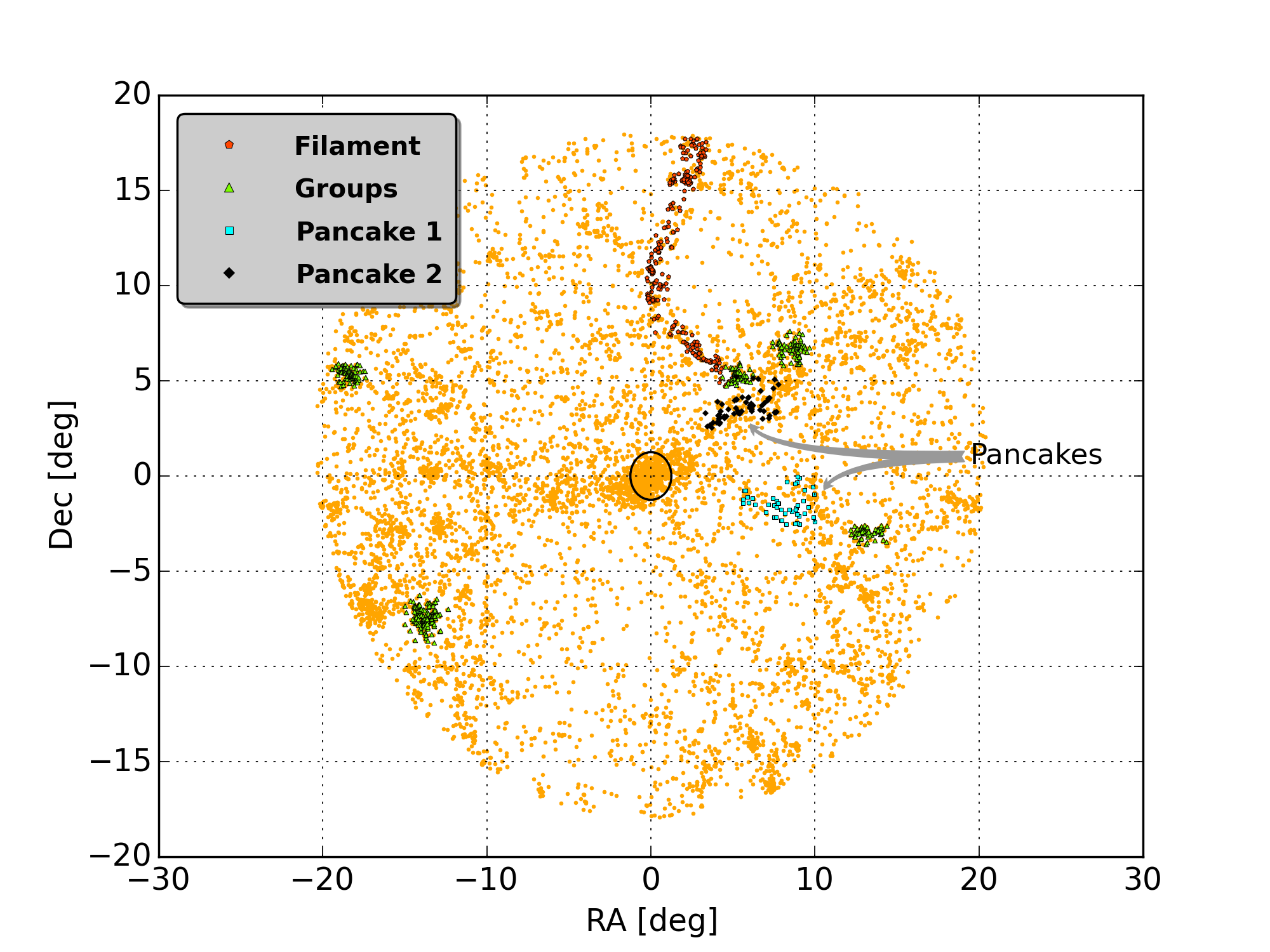}
\caption{Galaxies on the sky near the Coma cluster. Almost 9000 spectroscopically observed galaxies on the sky within an 18 degrees aperture. Only galaxies with line of sight velocities $\pm 4000$ km/s of $\langle v_{\rm Coma} \rangle = 6925$ km/s are included. The solid black circle indicates the virial radius of 2.4 Mpc. The two pancakes, shown in coloured symbols (black diamonds at 2 o'clock and blue squares at 3 o'clock), extend up to almost 17 Mpc from the centre of Coma, and the biggest is up to 5 Mpc wide. In red pentagons we highlight a filament, coming down from north. Filaments appears as 1-dimensional structures in this projection, as opposed to pancakes which are 2-dimensional on the sky. In green triangles we show four galaxy groups which clearly are localized in space \citep{Yang2007}.}
\label{fig:ComaRADec}
\end{figure}
\begin{figure}[t]
\vspace{-1cm}
\centering
\includegraphics[width=0.66\textwidth]{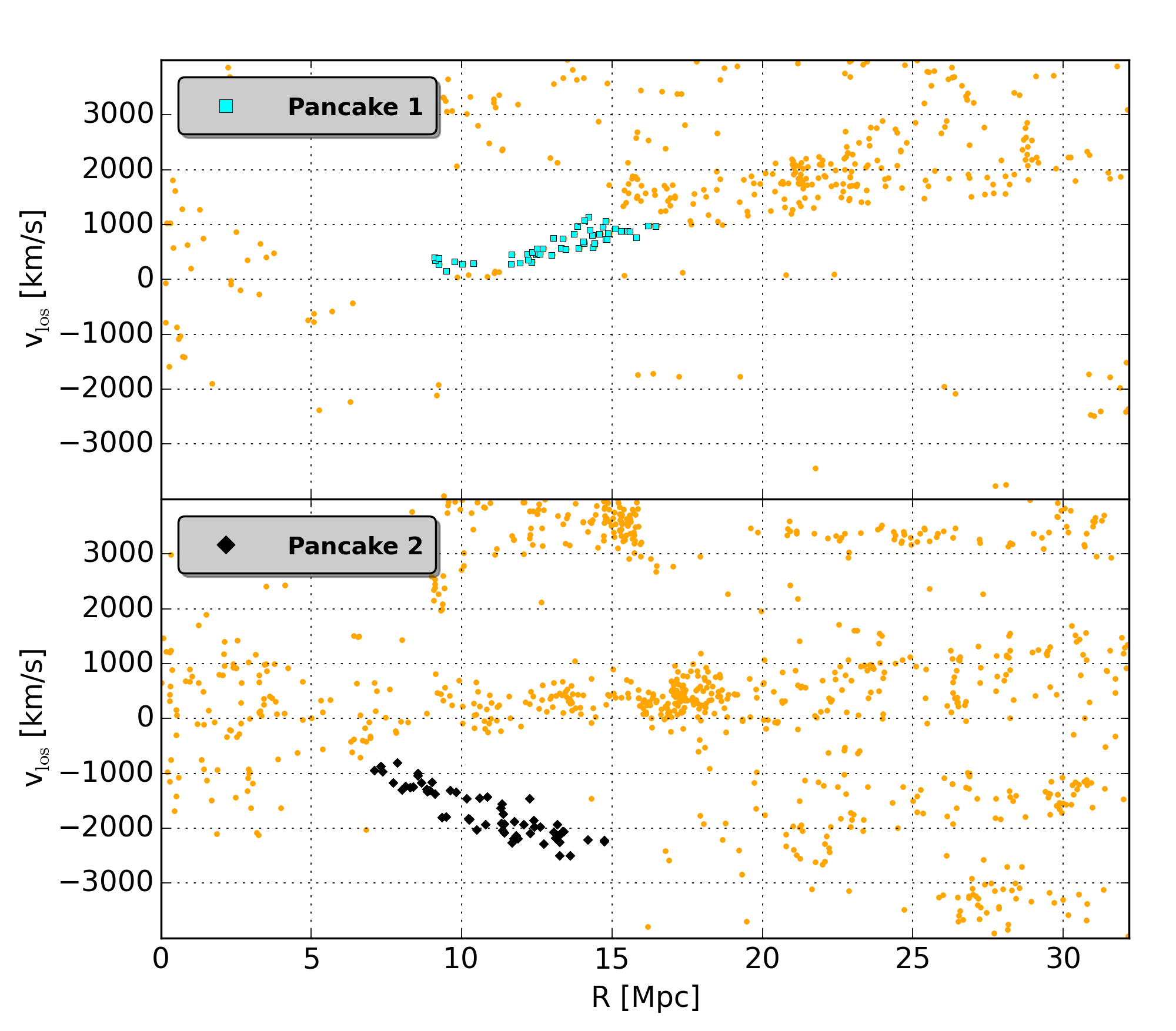}
\caption{\textbf{Top}: Line of sight velocity as a function of radius for the region (wedge containing 3/64 of the sky northeast of Coma) containing Pancake 1. The blue squares represent Pancake 1 and the orange circles the rest of the galaxies in this region. \textbf{Bottom}: The same as the top figure, but for the region (4/64 of the sky southeast of coma) containing Pancake 2, with the black diamonds representing Pancake 2. On both figures we see that the pancake stands out clearly from the surrounding area.}
\label{fig:duoRv}
\end{figure}
\begin{figure}[t]
%\vspace{-1cm}
\centering
\includegraphics[width=0.66\textwidth]{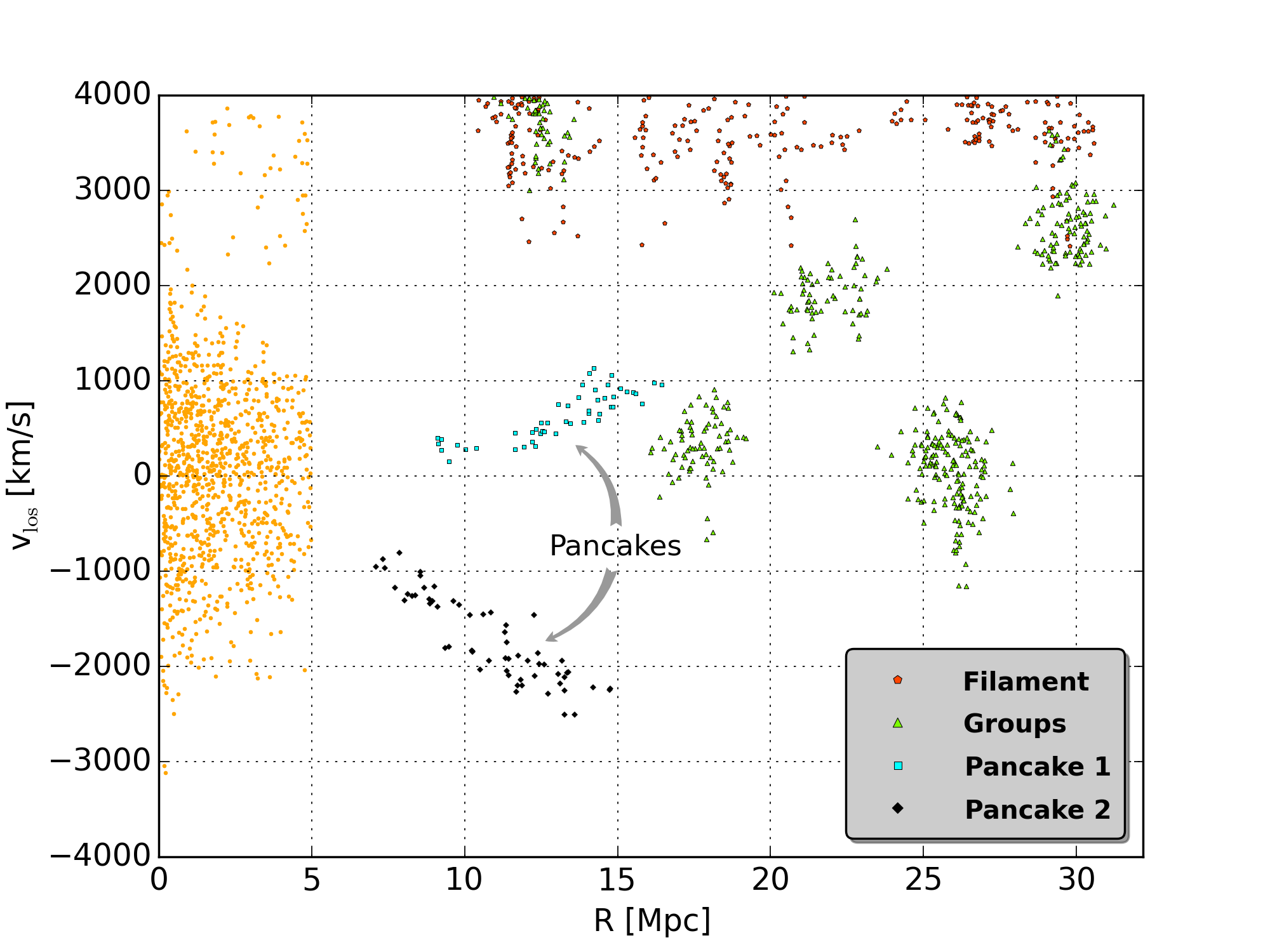}
\caption{Identification of pancakes in overdensities in velocity space. The orange dots show all the galaxies within a 5 Mpc radius of Coma. All other galaxies not belonging to one of the highlighted structures have been omitted. The two pancakes (blue squares and black diamonds) reveal themselves as inclined narrow collections of galaxies in this radius-velocity figure, exactly as cold pancakes near galaxy clusters are expected to. Pancake 1 (blue squares) points away from us, whereas Pancake 2 (black diamonds) sits between us and Coma. The long filament is shown in red pentagons (in the top of this figure), with no discernible angle $\theta$ (see Figure \ref{fig:projection} (right)), in a relatively narrow velocity interval (compared to the huge spatial extent), demonstrating that it really is a 1-dimensional structure. The five galaxy groups are shown as coloured triangles.}
\label{fig:ComaRv}
\end{figure}
In this section, we will present and discuss results obtained by using our method on spectroscopically observed galaxies near the Coma galaxy cluster. The data is from the Sloan Digital Sky Survey Data Release 7~\citep{Abazajian2009}. We select almost 9000 galaxies in an 18 degree aperture centered on Coma, where we take the galaxy NGC 4874~\citep{KentGunn1982} as the centre of the Coma cluster, within $\pm4000$ km/s of $<$v$_{\rm Coma}>$ = 6925 km/s.

Having calibrated our method from the numerical simulation, we look for pancakes in observational data.

We have identified two pancakes near the Coma galaxy cluster. These pancakes are the same as in \citet{Falco}. Pancake 1, however, is quite different from the one presented there. We optimized and froze our selection parameters using only the numerical simulation, which meant that only the low radius portion of the pancake from \citet{Falco} was selected. The high radius portion has much greater velocity dispersion and contains an overlapping galaxy group. Our method is only optimized to select overdensities in phase-space and did not include this portion. These pancakes are located at great distances from Coma, extending out from 7-8 Mpc from the centre and stretching across roughly 8 Mpc in length and up to 4-5 Mpc in width. These two pancakes, along with the largest galaxy groups and a filament near Coma can be seen in (RA,Dec) in Figure \ref{fig:ComaRADec}. From that figure we can see that the pancakes do not represent a significant overdensity when viewed on the sky.

Pancake 1, along with the background in the signal region, can be seen in (R,v$_{\rm los}$) in Figure \ref{fig:duoRv} (top) and a similar plot for Pancake 2 can be seen in Figure \ref{fig:duoRv} (bottom), from which we can see that the pancakes stand out as clear overdensities in (R,v$_{\rm los}$).

In Figure \ref{fig:ComaRv} we have plotted (R,v$_{\rm los}$) for pancakes 1 and 2, as well as the groups and filament indicated in Figure \ref{fig:ComaRADec}. It should be noted that most of these structures are not located in the same region on the sky. We see that the pancakes have a vastly different appearance in (R,v$_{\rm los}$) than the groups and the filament. The groups appear as high density, centralized structures and the filament is seen as an elongated one dimensional string of galaxies, whereas the pancakes are less dense, two dimensional structures, spread out over a greater region of the sky. The groups are from \citet{Yang2007} and the filament was found by eye as a spatial overdensity since it cannot be detected in projected phase-space using our method.

\section{Pancakes are cold structures}
\label{sec:Veldisp}
\begin{figure}
\centering
\begin{subfigure}
\centering
\includegraphics[width=0.75\textwidth]{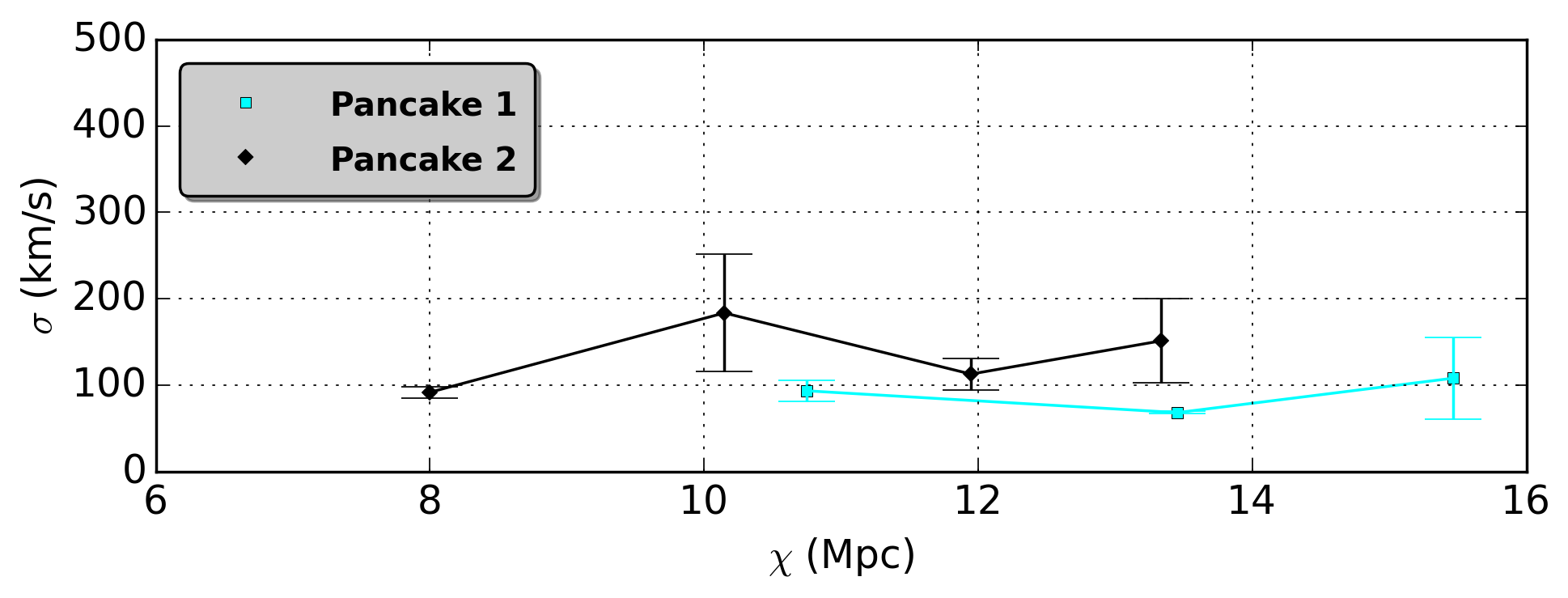}
\end{subfigure}\\
\begin{subfigure}
\centering
\includegraphics[width=0.75\textwidth]{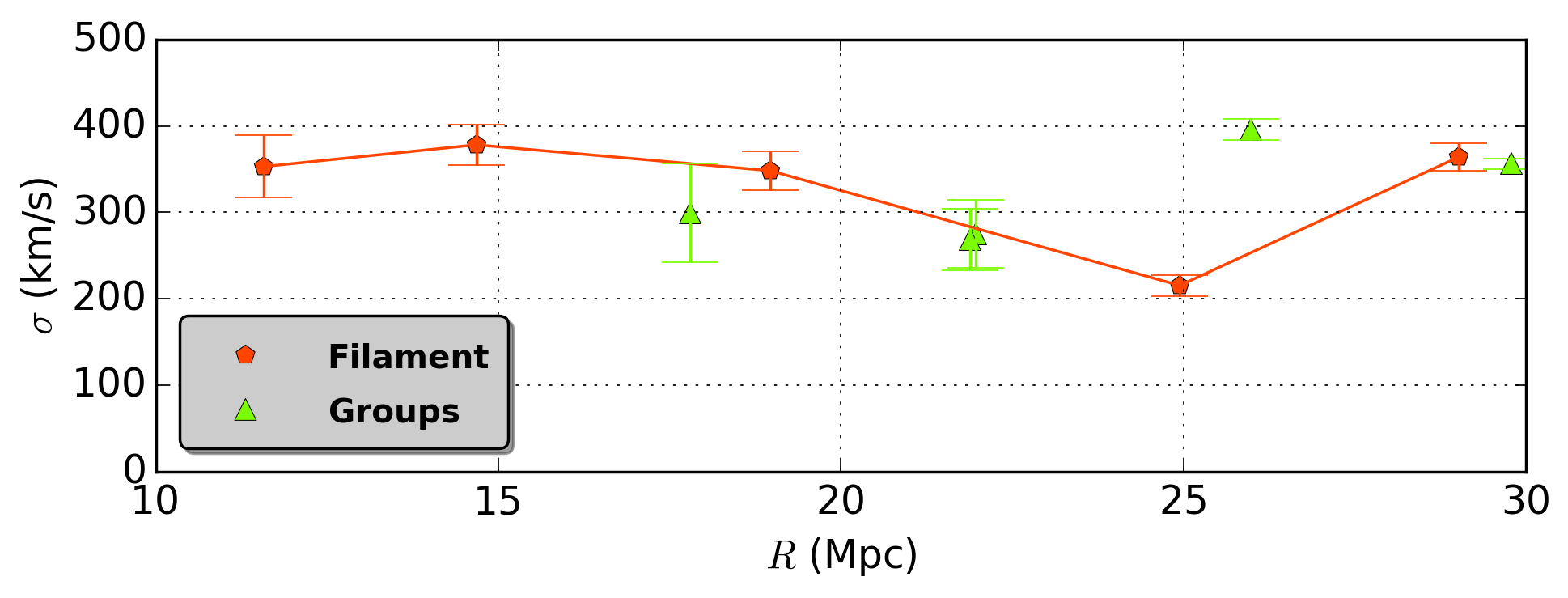}
\end{subfigure}\\
\caption{\textbf{Top}: The velocity dispersion perpendicular to the pancake, $\sigma$, as a function of the distance along the pancake, $\chi$. The aqua squares indicate Pancake 1 and the black diamonds Pancake 2. The figure shows fairly low velocity dispersion values of around 100 to 200 km/s. \textbf{Bottom}: Velocity dispersion in the v$_{\rm los}$ direction as a function of radius for the galaxy groups and the filament from Figure \ref{fig:ComaRADec}, where the green triangles are the galaxy groups indicated there and the red pentagons indicate the filament. In contrast to the top figure, this figure displays high velocity dispersion values mostly in the region of 300 to 400 km/s. Compared to the velocity dispersion of the galaxy groups and the filament, the velocity dispersions of the pancakes are much lower, which implies that the pancakes are cold structures.}
\label{fig:vdisp}
\end{figure}

\begin{figure}
\centering
\begin{subfigure}
\centering
\includegraphics[width=0.75\textwidth]{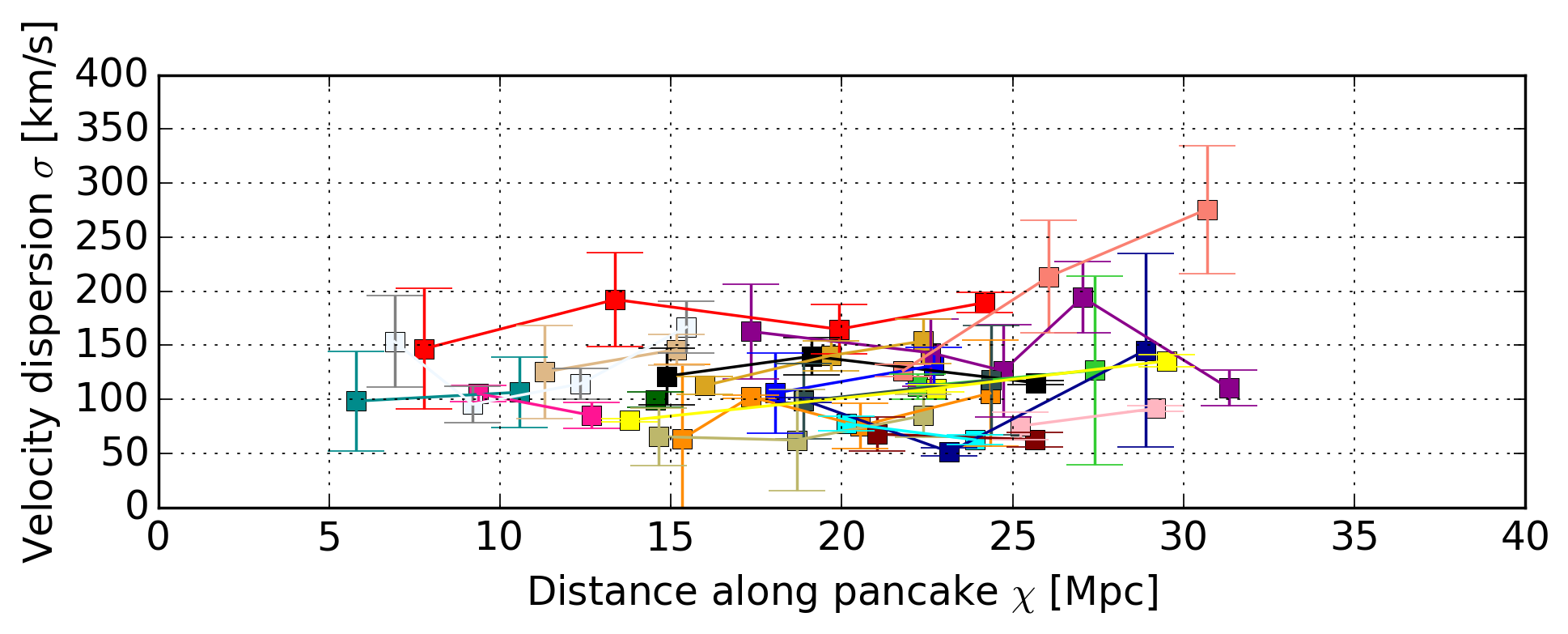}
\end{subfigure}\\
\begin{subfigure}
\centering
\includegraphics[width=0.75\textwidth]{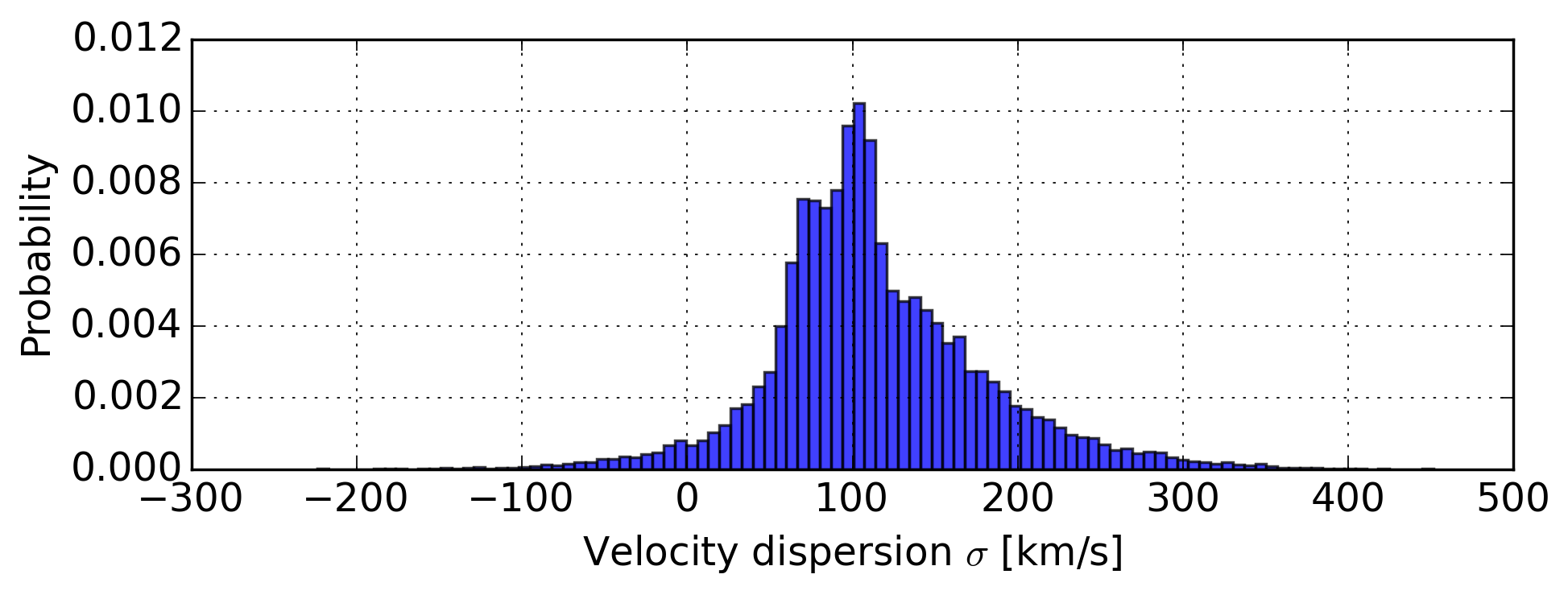}
\end{subfigure}\\
\caption{\textbf{Top}: The velocity dispersion perpendicular to the pancake, $\sigma$, as a function of the distance along the pancake, $\chi$, is displayed for the 20 pancakes identified in the numerical simulation. The figure shows fairly low velocity dispersion values of around 50 to 200 km/s. \textbf{Bottom}: Probability as a function of the velocity dispersion perpendicular to the pancake for the 20 numerical pancakes. We represented each pancake with a Gaussian distribution centred on the mean velocity dispersion value for a given pancake, which is the reason for negative velocity dispersion in the tail. The velocity dispersion distribution is centred at $103.5^{+66.9}_{-49.4}$ km/s, which is in agreement with the velocity dispersion values found for the two Coma pancakes.}
\label{fig:vdisp_num}
\end{figure}

An inherent quality of pancakes is that they have cold, well defined and contained flows away from a galaxy cluster (because of the Hubble flow). By calculating the velocity dispersion of a pancake, we can get a measure for the coldness and coherence of the pancake.

In order to measure the dispersion of the pancake, we fit the universal infall velocity, using the method outlined by ~\citep{Falco}, to the pancake galaxies in (R,v) space. Subtracting this fit, we get a scatter around zero, and we can measure the dispersion as function of distance along the pancake, $\chi$.\\

The velocity dispersion for the two Coma pancakes can be seen in Figure \ref{fig:vdisp} (top). In Figure \ref{fig:vdisp_num} (top) we show a similar velocity dispersion plot for a total of 20 pancakes identified in the numerical simulation. In Figure \ref{fig:vdisp_num} (bottom) we display the velocity dispersion probability distribution for these 20 numerical pancakes. The probability distribution was obtained by representing each pancake with a Gaussian distribution of 1000 data points centred on the mean velocity dispersion for each pancake, which is also the reason for the negative velocity dispersion values seen in the figure.

We see that the velocity dispersion for the two Coma pancakes is in good agreement with that found for the 20 numerical pancakes and their associated probability distribution.

Comparing the pancake velocity dispersions to that of the galaxy groups and filament near Coma (from Figure \ref{fig:ComaRADec}), which are shown in Figure \ref{fig:vdisp} (bottom), will indicate whether the pancakes are in fact cold, as we would expect.

We immediately see that the velocity dispersion values of the pancakes typically are one third to half that of the velocity dispersion of the galaxy groups and the filament, indicating that our pancakes indeed are cold and coherent.

\section{Pancake mass estimation}
\label{sec:Panmass}
The dispersion may have two independent origins. The first is related to the finite depth of the pancake, implying that the Hubble expansion across the pancake varies. This will be relevant only if the pancake is in the early stage of collapsing, and still far from an equilibrated state. The other is related to a "thermal" velocity of the pancake in equilibrium, which is the 2-dimensional equivalent to the 3-dimensional solution of the spherical Jeans equation, giving $G M_{\rm 3D} \approx 2 r \sigma^2$.

The low observed velocity dispersions of all the pancakes and the galaxy distribution of the numerical pancake indicate that the Hubble expansion contribution to the dispersion is likely to be minor.

We assume that the numerical pancakes and the observational pancakes are in similar states of equilibrium. So if we can ignore the Hubble expansion contribution to the dispersion, then the dispersion arises from the solution to the Jeans equation in two dimensions. Considering the moments of the collisionless Boltzmann equation for a disk-like structure one has 
\begin{align}
G M_{\rm pancake} = k_{\rm M} \sigma^2_{\rm M} \frac{A}{d},
\end{align}
where $A$ is the area on the sky, and $d$ is the half-light thickness of the pancake. $\sigma^2_{\rm M}$ is measured perpendicular to the pancake. For the 20 pancakes identified in the numerical simulation we know all of these values and as such we find the pre-factor to be $k_{\rm M} = 10^{-1.43 \pm 0.23}$. Using averages from the 20 numerical pancakes we can approximate the mass of the pancakes as
\begin{align}
M_{\rm pan} &= 4.2 \cdot 10^{15} M_{\astrosun} \nonumber \\
&\times \left(\frac{k_{\rm M}}{10^{-1.43}}\right)\left(\frac{\sigma_{\rm M}}{100 \rm km/s}\right)^2 \left(\frac{A}{110 \rm Mpc^2}\right) \left(\frac{2 \rm Mpc}{d}\right).
\label{eq:Mpan}
\end{align}

Observationally we can estimate both $\sigma^2$ and the area, however, the depth is more difficult. We use the 20 numerical pancakes to this end, wherefrom we can either use $d=2\left(\frac{\sigma}{100}\right)^2$Mpc or $d=2$Mpc. Both are fair approximations to the 20 numerical pancakes, and both are correct within a factor of two.

The areas on the sky are found to be approximately 37 Mpc$^2$ in projection for both Coma pancakes. Using the measured angle of orientation from observations ($\cos \alpha = 0.66$ and $0.35$) we get the actual area. We use $k_{\rm M}$ as found in the numerical simulation. Inserting these numbers, we find the total masses of the two Coma pancakes to be $2.1 \cdot 10^{15} M_{\astrosun}$ and $3.9 \cdot 10^{15}M_{\astrosun}$ respectively, using $d=2\left(\frac{\sigma}{100}\right)^2$Mpc (or $1.8 \cdot 10^{15} M_{\astrosun}$ and $7.2 \cdot 10^{15} M_{\astrosun}$ using $d=2$Mpc). The precision of this estimate is about a factor of 2 from the value of $k_{\rm M}$ and we estimate a factor of 2 for the depth of the pancakes. Additionally we estimate a factor of 2 from the velocity dispersion, because we cannot disentangle the velocity dispersion of the pancake from that of potentially embedded galaxy groups. The results are therefore accurate within a factor of 8.

\textbf{Stellar mass estimates.} Having identified the galaxies belonging to the pancakes, we can add the estimates of the stellar mass from \citep{Yang2007}. We find the stellar masses to be approximately $M_* = 1 \cdot 10^{12} M_{\astrosun}$ for Pancake 1 and $M_* = 5 \cdot 10^{11} M_{\astrosun}$ for Pancake 2.

This gives us an M$_*$/M$_{\rm pan}$ ratio of approximately $4 \cdot 10^{-4}$ and $5 \cdot 10^{-4}$ for Pancake 1 and $1 \cdot 10^{-4}$ and $7 \cdot 10^{-5}$ for Pancake 2, indicating that the pancakes are extremely dark matter dominated, in good agreement with them being in an early stage of evolution.
\section{Conclusion}
\citet{Falco} presented a novel method for identifying Zeldovich pancakes in observational data. We expanded and refined the method, and we examined the properties of two observational pancakes found near the Coma cluster using the method. This method differs from others in the literature, by identifying observational pancakes on relatively small scales of 10 Mpc.

We show that these structures are relatively cold, compared to typical groups of galaxies and filaments, with a velocity dispersion of around 100 km/s. These velocity dispersions are consistent with that of twenty pancakes identified in a cosmological simulation. We show that these simulation pancakes are only significantly collapsed or collapsing in one dimension. Additionally, we determine the stellar to total mass ratio of the observational pancakes to be $2 \cdot 10^{-4}$, within one order of magnitude.

In the near future we plan to fully automatize the pancake identification method and examine the environs of a large number of galaxy clusters. This should provide a statistically significant sample of observational pancakes, on the scale of 10 Mpc, for further study.

\section*{Acknowledgements}
The authors thank Radoslaw Wojtak and Martin Sparre for numerous discussions and physical insights, Johan Samsing and Jesús Zavala for comments and helpful discussions, and Stefan Gottloeber, who kindly agreed for one of the CLUES simulations (http://www.clues-project.org/simulations.html) to be used in the paper. The simulation has been performed at the Leibniz Rechenzentrum (LRZ) Munich. The Dark Cosmology Centre is funded by the Danish National Research Foundation. MF acknowledges partial support from the INFN grant In-dark and from the grant Progetti di Ateneo / CSP TO Call2 2012 0011 “Marco Polo” of the University of Torino.

\bibliographystyle{JHEPb}

\end{document}